\documentclass[reprint,
groupedaddress,
showpacs,preprintnumbers,
 amsmath,amssymb,
 aps,
prb,
longbibliography
]{revtex4-1}
\usepackage{graphicx}
\usepackage{dcolumn}
\usepackage{bm}
\usepackage{mathrsfs}
\usepackage{hyperref}
\usepackage{paralist}

\begin{document}

\title{Influence of the impurity scattering on charge transport in
	unconventional superconductor junctions 
}
\author{Bo Lu$^{1}$, Pablo Burset$^{1}$, Yasunari Tanuma$^{2}$, Alexander A. Golubov$^{3,6}$, Yasuhiro Asano$^{4,5,6}$, Yukio Tanaka$^{1,6}$}

\affiliation{$^1$~Department of Applied Physics, Nagoya University, Nagoya 464-8603, Japan\\
$^2$~Graduate School of Engineering Science, Akita University, Akita 010-8502, Japan\\
$^3$~Faculty of Science and Technology and MESA+ Institute for Nanotechnology,
University of Twente, 7500 AE Enschede, The Netherlands\\
$^4$~Department of Applied Physics, Hokkaido University, Sapporo 060-8628, Japan\\
$^5$~Center of Topological Science and Technology, Hokkaido University, Sapporo 060-8628, Japan\\
$^6$~Moscow Institute of Physics and Technology, Dolgoprudny, Moscow 141700, Russia
}
\date{\today}

\begin{abstract}
We study the influence of non-magnetic impurity scatterings on the tunneling conductance
of a junction consisting of a normal metal and a disordered unconventional superconductor
by solving the quasiclassical Eilenberger equation self-consistently.
We find that the impurity scatterings in both the Born and unitary
limits affect the formation of the Andreev bound states and modify strongly the tunneling spectra around zero bias.
Our results are interpreted well by the appearance of odd-frequency Cooper pairs near the interface
and by the divergent behavior of the impurity self-energy.
The present
paper provides a useful tool to identify the pairing symmetry of unconventional superconductors in
experiments.
\end{abstract}

\maketitle


\section{Introduction}
The effects of impurity scatterings on superconducting phenomena are a key issue 
in the field of superconductivity. While the BCS s-wave pairing state is
robust against nonmagnetic impurities~\cite{God}, unconventional pairing states with other
symmetries are usually rather fragile \cite%
{Larkin,Balian,Millis88,Sigrist91,Radtke93}. 
The impurity scatterings modify
transport and thermodynamic properties of unconventional superconductors,
which has a crucial impact on the identification of the pairing symmetry, e.g., 
in $p$-wave superconductors
~\cite{Buchholtz1981,MakiEPL2000,BalaPRB2000}.
Specifically, the tunneling spectroscopy is an important
experimental tool to identify an unconventional superconductor.
The formation of the Andreev bound states at the surface of a superconductor 
is attributed to the anisotropy in the pairing states~\cite%
{Buchholtz1981,bruder,Hu94,KashiwayaRev} displaying a pronounced zero-bias
conductance peak (ZBCP) in the tunneling conductance. In some cases, such
bound states have a topological origin due to the bulk-edge correspondence~\cite%
{Schnyder2008,sato1,sato2,Hasan10,Ryu10,Hasan11,ZhangRev,AliceaRev,TanakaRev2012}.
Therefore, the ZBCP in the tunneling
conductance strongly suggests unconventional pairing symmetries of a superconductor.

Theoretically, the tunneling spectroscopy is formulated as a differential 
conductance in a normal metal/superconductor junction. 
Although effects of the potential disorder on the conductance have been 
studied in a number of papers, many of them focus on the disorder in the normal 
metal~\cite{diffusive1,Tanaka04,Tanaka05,Golubov97}. 
Surprisingly, only few attempts have been made at the effects of disorder in the 
superconductor. 
Exceptional examples may be the studies about the proximity effect\cite{Lofwander04} and the ZBCP in high-Tc cuprate junctions. It is found that the brodened ZBCP can be explained by the impurity scatterings in the superconductor or the surface roughness\cite%
{bruder1998,Kupriyanov,TanumaImpurity2001,Asano02}. 
Although $p$-wave or chiral $p$-wave superconductors are a recent topic of interest, 
effects of disorder in such superconductors have not yet been systematically studied.  
For instance, we have never known if the dome-like subgap conductance
in chiral $p$-wave superconductors\cite{Yamashiro97,Honerkamp,yamashira99} is robust or not in the presence of impurities.
Similarly, chiral $d$- and chiral $f$-wave symmetries have been recently
suggested to explain the absence of spontaneous edge currents in 
Sr$_{2}$RuO$_{4}$ \cite{Huang14,Tada15}, where the sensitivity of the chiral edge current to 
potential disorder depends on chiral pairing symmetries\cite{Suzuki}.
Several theoretical papers propose that doped graphene at a van Hove
singular point may develop a chiral $d$-wave pairing state \cite%
{Schaffer14,Nandkishore12,Wang,Kiesel12}, which may be sensitive to weak
impurity potentials treated within the Born limit. 
Furthermore, there are increasing evidences that suggest the strong impurity
potential in Sr$_{2}$%
RuO$_{4}$\cite{Maeno1994,Kallin1999,Maeno2003,Maeno2013}. 
Theoretically, such strong impurities scatterings should be described by 
the unitary limit\cite%
{Miyake1999,impuritySrRUO}. 
As a result, a plethora of novel systems require
a detailed analysis of the impurity effects on unconventional pairings in
both weak and strong impurity limits.

In this article, we present a self-consistent calculation of the tunneling
spectra in a normal-metal/disordered-superconductor junction with flat
or chiral surface band states. For the former case, occurring in nodal
superconductors, we find that the symmetry of the emergent odd-frequency
pairing states \cite{Berezinskii,Kirkpatrick,Balatsky92,Bergeret05,Golubov97} plays a pivotal role in determining the
evolution of the ZBCP under the impurity scatterings. 
As for chiral superconductors, our results show that the influence of the impurity
scattering on the zero bias conductance in the unitary limit is more
pronounced than that in the Born limit. It is important to emphasize that the
disorder strength we consider is still below the diffusive (or Usadel) limit
\cite{Usadel}. In that case, the strong pair-breaking effect would make the
unconventional superconductor gapless.

The remainder of this paper is organized as follows. In section II, we
describe our model and derive the essential formulas. In section III, we
discuss impurity effects in junctions with nodal superconductors. The order
parameters and tunneling spectroscopy are studied. In section IV, we show
our result for chiral superconductors. Some concluding remarks are given in
section V.

\section{model and formalism}

\begin{figure}[tbph]
	\includegraphics[width = 75 mm]{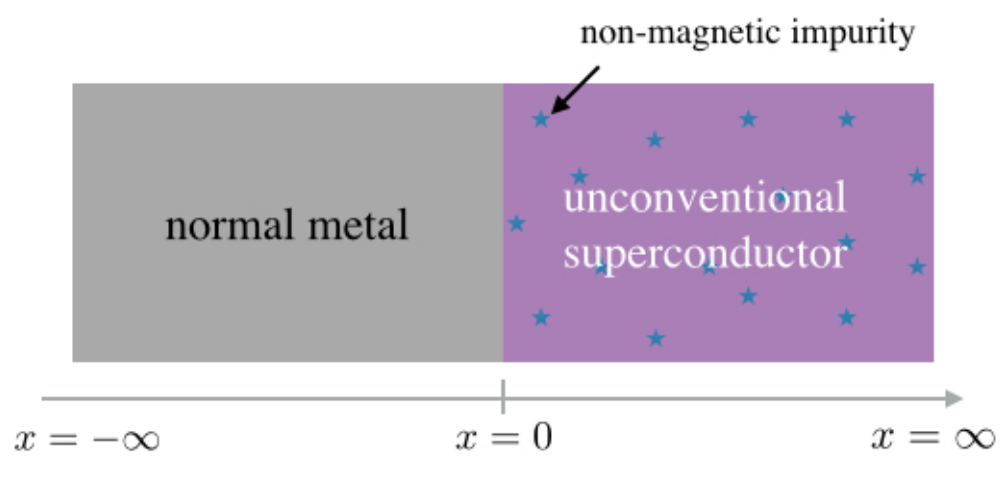}
	\caption{(Color online) Schematic illustration of a normal
		metal/unconventional superconductor junction with a flat interface at $x=0$.
		We assume impurities in superconducting side. }
	\label{fig0}
\end{figure}

Let us consider a two-dimensional normal metal-superconductor junction with
a flat interface at $x=0$ as shown in Fig.\ref{fig0}.
The interface barrier potential is described by a delta function $V\left(
x\right) =H\delta \left( x\right) $, with $H$ the potential strength. The
normal metal $\left( x<0\right) $ is assumed to be clean while the
superconductor $\left( x>0\right) $ has a uniform impurity distribution. We
denote the quasiclassical Green's function \cite%
{SereneRainer,Zaitsev1984,Sauls1988,Nagai1989} by $\hat{g}^{\alpha \alpha
}\left( i\omega _{n},x,\theta _{\alpha }\right) $, with Matsubara frequency $%
\omega _{n}=(2n+1)\pi T$, temperature $T$, and an integer $n$. $\theta _{\pm
}$ is given by $\theta _{+}\equiv \theta \in \left[ -\pi /2,\pi /2\right] $
and $\theta _{-}=\pi -\theta $. We focus on the spin-degenerate system where
$\hat{g}^{\alpha \alpha }\left( i\omega _{n},x,\theta _{\alpha }\right) $ is
a $2\times 2$ matrix in particle-hole space satisfying Eilenberger equation%
\cite{Eilenberger}
\begin{equation}
iv_{f}\cos \theta \partial _{x}\hat{g}^{\alpha \alpha }+\alpha \left[
i\omega _{n}\hat{\tau}_{3}-\hat{\Delta}(\theta _{\alpha },x)\hat{\tau}_{3}-%
\hat{a}\left( x\right) ,\hat{g}^{\alpha \alpha }\right] =0,
\end{equation}%
accompanied by the normalization condition $\left( \hat{g}^{\alpha \alpha
}\right) ^{2}=-1$. Here $v_{f}$ is the quasi-particle Fermi velocity and $%
\hat{\tau}_{i}$ are Pauli matrices. $\hat{\Delta}(\theta _{\alpha },x)$ and $%
\hat{a}\left( x\right) $ are the superconducting order parameter and the
self-energy induced by impurities, respectively. In the normal metal region,
$\hat{\Delta}(\theta _{\alpha },x<0)=\hat{a}\left( x<0\right) =0$. We
consider superconductors with different kinds of pairing symmetries: $d_{xy}$%
, $p_{x}$, $p_{y}$, and chiral states, so it is convenient to express the
order parameter in the superconducting side $x>0$ as
\begin{equation}
\hat{\Delta}(\theta _{\alpha },x)=-\Delta _{1}(x,\theta _{\alpha })\hat{\tau}%
_{2}+\Delta _{2}(x,\theta _{\alpha })\hat{\tau}_{1}.
\end{equation}%
For $d_{xy}$-, $p_{x}$-, and $p_{y}$- waves, $\Delta _{1}(x,\theta _{\alpha
})=0$ and the gap equation $\Delta _{2}(x,\theta _{\alpha })=\Delta \left(
x\right) \chi \left( \theta _{\alpha }\right) $ is given by using the
quasi-classical Green's function as%
\begin{equation*}
	\Delta \left( x\right) =\frac{\displaystyle2T\sum\limits_{0<\omega
			_{m}<\omega _{c},\alpha }\int_{-\frac{\pi }{2}}^{\frac{\pi }{2}}d\theta
		^{\prime }\chi \left( \theta _{\alpha }^{\prime }\right) \left[ \hat{g}%
		^{\alpha \alpha }\left( i\omega _{m},\theta _{\alpha }^{\prime },x\right) %
		\right] _{12}}{\displaystyle\ln \frac{T}{T_{c}}+\sum_{0<m<\frac{\omega _{c}}{%
				2\pi T}}\frac{1}{m-1/2}},
\end{equation*}%
where $\chi \left( \theta _{\alpha }\right) $ is chosen as $\cos \theta
_{\alpha }$ ($p_{x}$-wave), $\sin \theta _{\alpha }$ ($p_{y}$-wave), and $%
\sin 2\theta _{\alpha }$ ($d_{xy}$-wave). $T_{c}$ and $\omega _{c}$ are the
critical temperature of the bulk superconductor and the Debye frequency,
respectively. For chiral wave states, $\Delta _{1}(x,\theta _{\alpha
})=\Delta _{im}\left( x\right) \chi _{im}\left( \theta _{\alpha }\right) $
and $\Delta _{2}(x,\theta _{\alpha })=\Delta _{re}\left( x\right) \chi
_{re}\left( \theta _{\alpha }\right) $ are given by%
\begin{equation*}
\Delta_{re} \left( x\right) =
\frac{\displaystyle \sum\limits_{0<\omega_{m}<\omega_{c},\alpha }
	\int_{-\frac{\pi }{2}}^{\frac{\pi }{2}%
	}d\theta ^{\prime }\chi_{re} \left( \theta _{\alpha }^{\prime }\right)
	\left[
	\hat{g}^{\alpha \alpha }
	\left( i\omega _{m},\theta _{\alpha }^{\prime},x\right)
	\right] _{12}}{
	\displaystyle
	\frac{1}{2T}
	\left(
	\displaystyle
	\ln \frac{T}{T_{c}}+\sum_{0<m<\frac{\omega _{c}}{%
			2\pi T}}\frac{1}{m-1/2}
	\right)},
\end{equation*}
and%
\begin{equation*}
\Delta_{im} \left( x\right) =
\frac{\displaystyle \sum\limits_{0<\omega_{m}<\omega _{c},\alpha }
	\int_{-\frac{\pi }{2}}^{\frac{\pi }{2}%
	}d\theta ^{\prime }\chi_{im} \left( \theta _{\alpha }^{\prime }\right)
	\left[
	\hat{g}^{\alpha \alpha }
	\left( i\omega _{m},\theta _{\alpha }^{\prime},x\right)
	\right] _{12}}{
	\displaystyle
	\frac{i}{2T}
	\left(
	\displaystyle
	\ln \frac{T}{T_{c}}+\sum_{0<m<\frac{\omega _{c}}{%
			2\pi T}}\frac{1}{m-1/2}
	\right)},
\end{equation*}
with%
\begin{align*}
	\chi _{re}\left( \theta _{\alpha }\right) ={}& \cos \left( \lambda \theta
	_{\alpha }\right) , \\
	\chi _{im}\left( \theta _{\alpha }\right) ={}& \sin \left( \lambda \theta
	_{\alpha }\right) .
\end{align*}%
We only consider the cases with chiral $p$-wave ($\lambda =1$), chiral $d$%
-wave ($\lambda =2$), and chiral $f$-wave ($\lambda =3$).

The self-energy in the superconducting side can be decomposed as $\hat{a}%
=a_{1}\hat{\tau}_{1}+a_{2}\hat{\tau}_{2}+a_{3}\hat{\tau}_{3}$ and is
connected to the quasiclassical Green's function by
\begin{equation}
a_{j}\left( i\omega _{m},x\right) =-\frac{1}{2\tau }\frac{\frac{1}{1-\sigma }%
	\sum_{\alpha }\left\langle g_{j}^{\alpha \alpha }\left( i\omega
	_{m},x\right) \right\rangle _{\theta }}{1-\frac{\sigma }{1-\sigma }%
	\sum_{i,\alpha }\left\langle g_{i}^{\alpha \alpha }\left( i\omega
	_{m},x\right) \right\rangle _{\theta }^{2}},
\end{equation}%
with $\hat{g}^{\alpha \alpha }=\sum_{i}g_{i}^{\alpha \alpha }\hat{\tau}_{i}$%
. Here, $1/\tau $ and $\sigma $ denote the normal scattering rate and
strength of a simple impurity potential, respectively. In the Born limit one
finds that $\sigma \rightarrow 0$, while in the unitary limit $\sigma
\rightarrow 1$, resulting in%
\begin{equation}
\hat{a}=\left\{
\begin{array}{cc}
\displaystyle-\frac{1}{2\tau }\sum_{\alpha }\left\langle \hat{g}^{\alpha
	\alpha }\right\rangle  & \mbox{(Born limit)} \\
\displaystyle\frac{1}{2\tau }\left[ \sum_{\alpha }\left\langle \hat{g}%
^{\alpha \alpha }\right\rangle \right] ^{-1} & \mbox{(unitary limit)}%
\end{array}%
\right.   \label{eqself}
\end{equation}%
In the superconducting region, the Riccati parameterization\cite%
{RiccatiJapan1,RiccatiWestern,RiccatiEschrig} is%
\begin{equation}
\hat{g}^{++}=\frac{i}{1-\mathcal{G}_{+}^{S}\mathcal{F}_{+}^{S}}\left[
\begin{array}{cc}
1+\mathcal{G}_{+}^{S}\mathcal{F}_{+}^{S} & 2i\mathcal{F}_{+}^{S} \\
2i\mathcal{G}_{+}^{S} & -1-\mathcal{G}_{+}^{S}\mathcal{F}_{+}^{S}%
\end{array}%
\right] ,
\end{equation}%
\begin{equation}
\hat{g}^{--}=\frac{i}{\mathcal{G}_{-}^{S}\mathcal{F}_{-}^{S}-1}\left[
\begin{array}{cc}
1+\mathcal{G}_{-}^{S}\mathcal{F}_{-}^{S} & 2i\mathcal{F}_{-}^{S} \\
2i\mathcal{G}_{-}^{S} & -1-\mathcal{G}_{-}^{S}\mathcal{F}_{-}^{S}%
\end{array}%
\right] .
\end{equation}%
The Riccati parameters $\mathcal{G}_{\pm }^{S}$ , $\mathcal{F}_{\pm }^{S}$
obey the equations
\begin{subequations}
	\begin{alignat}{4}
		& v_{f}\cos \theta \partial _{x}\mathcal{G}_{+}^{S}=2\tilde{\omega}_{n}%
		\mathcal{G}_{+}^{S}-\Lambda _{2+}+\Lambda _{1+}\left( \mathcal{G}%
		_{+}^{S}\right) ^{2},\newline
		& & & & & & \\
		& v_{f}\cos \theta \partial _{x}\mathcal{F}_{+}^{S}=-2\tilde{\omega}_{n}%
		\mathcal{F}_{+}^{S}-\Lambda _{1+}+\Lambda _{2+}\left( \mathcal{F}%
		_{+}^{S}\right) ^{2}, & & & & & & \\
		& v_{f}\cos \theta \partial _{x}\mathcal{G}_{-}^{S}=-2\tilde{\omega}_{n}%
		\mathcal{G}_{-}^{S}+\Lambda _{2-}-\Lambda _{1-}\left( \mathcal{G}%
		_{-}^{S}\right) ^{2},\newline
		& & & & & & \\
		& v_{f}\cos \theta \partial _{x}\mathcal{F}_{-}^{S}=2\tilde{\omega}_{n}%
		\mathcal{F}_{-}^{S}+\Lambda _{1-}-\Lambda _{2-}\left( \mathcal{F}%
		_{-}^{S}\right) ^{2}, & & & & & &
	\end{alignat}%
	with the definitions
\end{subequations}
\begin{eqnarray*}
	\tilde{\omega}_{n} &=&\omega _{n}+ia_{3}, \\
	\Lambda _{1\pm } &=&i\Delta _{1\pm }+\Delta _{2\pm }-a_{1}+ia_{2}, \\
	\Lambda _{2\pm } &=&-i\Delta _{1\pm }+\Delta _{2\pm }+a_{1}+ia_{2}.
\end{eqnarray*}%
At the interface $x=0$, we have the boundary conditions\cite{RiccatiJapan1,RiccatiEschrig,Fogel2000,Zhao}%
\begin{equation}
{\cal F}^{S}_{+}(x=0) \to
\begin{cases}
\displaystyle
\frac{R}{{\cal G}^{S}_{-}(x=0)}, & \omega_{n}>0,
\\
\displaystyle
\frac{1}{R{\cal G}^{S}_{-}(x=0)}, & \omega_{n}<0,
\end{cases}
\end{equation}
\begin{equation}
{\cal F}^{S}_{-}(x=0) \to
\begin{cases}
\displaystyle
\frac{1}{R{\cal G}^{S}_{+}(x=0)}, & \omega_{n}>0,
\\
\displaystyle
\frac{R}{{\cal G}^{S}_{+}(x=0)}, & \omega_{n}<0,
\end{cases}
\end{equation}
with $R=Z^{2}/\left[ Z^{2}+4\cos ^{2}\theta \right] $ , $Z=2mH/k_{f}$, m the
electron mass, and $k_{f}$ the Fermi momentum. In this work, we only
consider the low transmitting case with fixed $Z=5$. 
The local density of states (LDOS) $\rho _{S}$ is given by%
\begin{equation}
\rho _{S}\left( x\right) =\frac{1}{\pi }\text{Im}\sum\limits_{\alpha }\int_{-%
	\frac{\pi }{2}}^{\frac{\pi }{2}}d\theta ^{\prime }\hat{g}^{\alpha \alpha
}\left( i\omega _{m}\rightarrow E+i\delta ,\theta _{\alpha }^{\prime
},x\right) _{11}
\end{equation}%
To obtain the conductance, we write down the wave functions in terms of the
Riccati parameter $\mathcal{\bar{G}}_{\pm }^{S}=\mathcal{G}_{\pm }^{S}\left(
E,\theta _{\pm },x=0\right) $ at the interface, which are
\begin{equation*}
	\left\{
	\begin{array}{l}
		\text{N: }\left[
		\begin{array}{c}
			1 \\
			a%
		\end{array}%
		\right] e^{ik_{f}\cos \theta x}+\left[
		\begin{array}{c}
			b \\
			0%
		\end{array}%
		\right] e^{-ik_{f}\cos \theta x} \\
		\text{S: }c\left[
		\begin{array}{c}
			1 \\
			i\mathcal{\bar{G}}_{+}^{S}%
		\end{array}%
		\right] e^{ik_{f}\cos \theta x}+d\left[
		\begin{array}{c}
			-i\left[ \mathcal{\bar{G}}_{-}^{S}\right] ^{-1} \\
			1%
		\end{array}%
		\right] e^{-ik_{f}\cos \theta x}%
	\end{array}%
	\right. ,
\end{equation*}%
with $a$, $b$ ($c$, $d$) reflection (transmission) amplitudes. Using BTK
theory\cite{BTK}, we obtain
the normalized differential conductance
\begin{equation*}
	\sigma _{t}=\int_{-\frac{\pi }{2}}^{\frac{\pi }{2}}d\theta \cos \theta
	\sigma _{\theta }/\sigma _{n},
\end{equation*}%
where $\sigma _{\theta }=1-\left\vert b\right\vert ^{2}+\left\vert
a\right\vert ^{2}$ reads\cite{TanakaD95,KashiwayaRev}%
\begin{equation}
\sigma _{\theta }=\sigma _{N}\frac{\left\vert \mathcal{\bar{G}}%
	_{-}^{S}\right\vert ^{2}+\sigma _{N}\left\vert \mathcal{\bar{G}}_{+}^{S}%
	\mathcal{\bar{G}}_{-}^{S}\right\vert ^{2}+\left( \sigma _{N}-1\right)
	\left\vert \mathcal{\bar{G}}_{+}^{S}\right\vert ^{2}}{\left\vert \mathcal{%
		\bar{G}}_{-}^{S}+\left( \sigma _{N}-1\right) \mathcal{\bar{G}}%
	_{+}^{S}\right\vert ^{2}},
\end{equation}%
with $\sigma _{N}=4\cos ^{2}\theta /\left[ Z^{2}+4\cos ^{2}\theta \right] $
and $\sigma _{n}=\int_{-\frac{\pi }{2}}^{\frac{\pi }{2}}d\theta \cos \theta
\sigma _{N}$ the conductance in normal states. In the numerical
calculations, we take the temperature $T=0.05T_{c}$. For convenience, we do
not consider thermodynamic phenomena and assume the Debye frequency $\omega
_{c}=2\pi $ for all the considered pairing states.

\begin{figure}[tbph]
	\includegraphics[width = 80 mm]{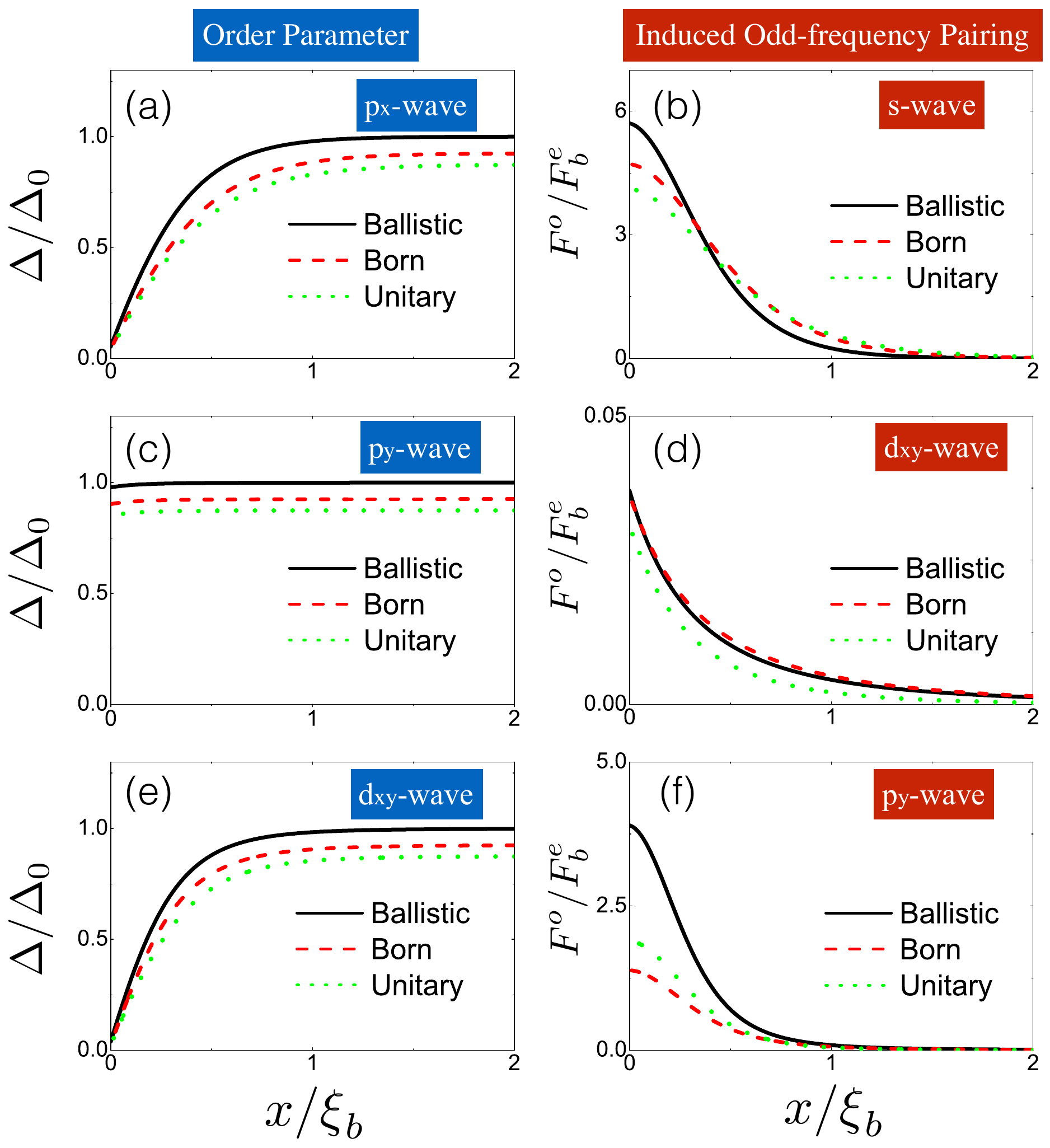}
	\caption{(Color online) Order parameter amplitude (left) and induced
		odd-frequency pairing amplitude (right) near the interface between a
		superconductor and a normal metal. From top to bottom, (a)(b) $p_{x}$-wave,
		(c)(d) $p_{y}$-wave, and (e)(f) $d_{xy}$ -wave.}
	\label{fig1}
\end{figure}

\section{Nodal superconductors with flat surface bands}

\subsection{Order parameter and odd-frequency}

First, we show the spacial variation of the order parameter amplitude in the
left panels of Fig.\ref{fig1}. Three regimes are considered: ballistic,
Born, and unitary limit. In the ballistic case, the relaxation time becomes
infinite so we set $1/\tau =0$. For Born and unitary limit, it is given by $%
1/\tau =0.2\Delta _{b}$, where $\Delta _{b}$ is the bulk value of the order
parameter (in ballistic case, the value of the order parameter $\Delta _{b}$
is denoted by $\Delta _{0}$). We define the superconducting coherence length
as $\xi _{b}=v_{f}/\Delta _{b}$. We can see that for $p_{x}$- and $d_{xy}$%
-wave cases, the order parameters with or without impurities strongly
decrease when they approach the interface. However, for the $p_{y}$ case,
the order parameter is almost invariant in space. For the following
discussion of conductance spectra in this subsection, we plot the
odd-frequency pairing states in the right panels of Fig.\ref{fig1}. The
anomalous Green's function $\bar{F}=\sum_{\alpha }\left[ \hat{g}^{\alpha
	\alpha }\right] _{12}$ can be decomposed into an even-frequency component $%
\bar{F}^{e}\left( i\omega _{n},\theta \right) =\bar{F}^{e}\left( -i\omega
_{n},\theta \right) $ and an odd-frequency one $\bar{F}^{o}\left( i\omega
_{n},\theta \right) =-\bar{F}^{o}\left( -i\omega _{n},\theta \right) $. The
Fourier transform of $\bar{F}^{o\left( e\right) }$ is given by\cite{TanakaPRL07,Eschrig2007,Tanuma07}%
\begin{equation}
\bar{F}^{o\left( e\right) }\left( i\omega _{n},\theta \right)
=\sum_{m}f_{c,m}^{o\left( e\right) }\cos \left( m\theta \right)
+f_{s,m}^{o\left( e\right) }\sin \left( m\theta \right) .
\end{equation}%
We denote by $F^{o}$ the dominant odd-frequency component $f_{c,m}^{o}$ or $f_{s,m}^{o}$
and normalize it by the bulk value of the even-frequency component $%
F_{b}^{e}=f_{c,m}^{e}$ (with $m=1,2$ for $p_{x}$-, $d_{xy}$-wave,
respectively) or $F_{b}^{e}=f_{s,1}^{e}$ ($p_{y}$-wave). It can be seen from
the results that as the order parameter decreases in $p_{x}$- and $d_{xy}$%
-wave cases, a large value of odd-frequency component is generated near the
interface which determines the shape of Cooper pairs to be $s$- and $p_{y}$%
-wave, respectively\cite{TanakaPRL07}. One can compare the magnitudes of odd-frequency
components in the presence of impurities with the ballistic case to find
that the generated $s$-wave odd-frequency magnitude is not sensitive to the impurity scattering which is consistent with the Anderson theorem\cite{God}. In the $d_{xy}$-wave case, the non-$s$-wave odd frequency component
is strongly suppressed as the impurity scattering is introduced. For $p_{y}$%
-wave case, the odd-frequency component is very small while the dominant
pairing near the surface is even-frequency.
\begin{figure}[tbph]
	\includegraphics[width = 75 mm]{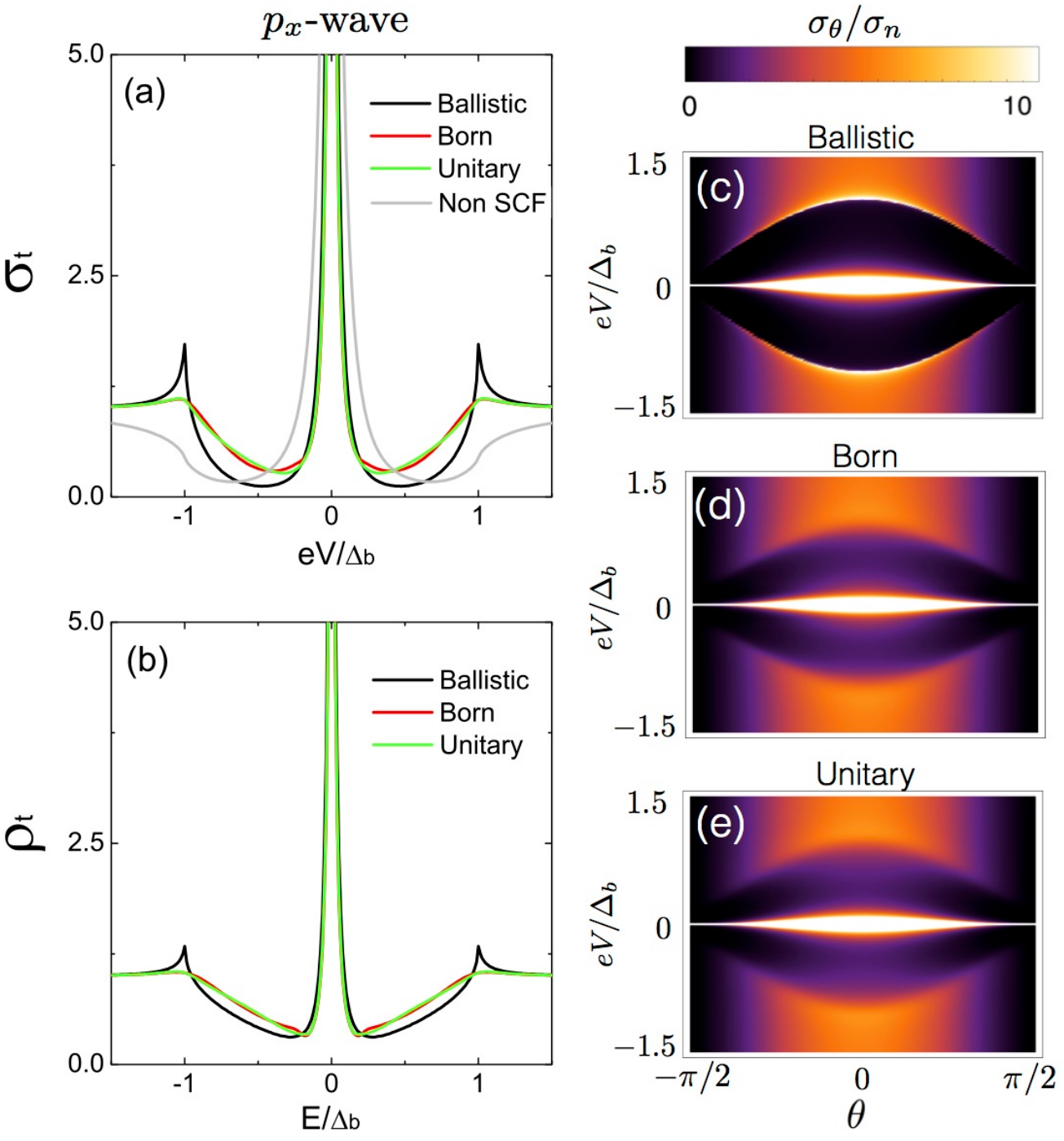}
	\caption{(Color online) $P_{x}$-wave: (a) conductance and (b)
		local density of states at the normal metal-superconductor interface.
		Angle-resolved conductance for (c) ballistic, (d) Born, and (e) unitary limit.
	}
	\label{fig2}
\end{figure}
\begin{figure}[tbph]
	\includegraphics[width = 75 mm]{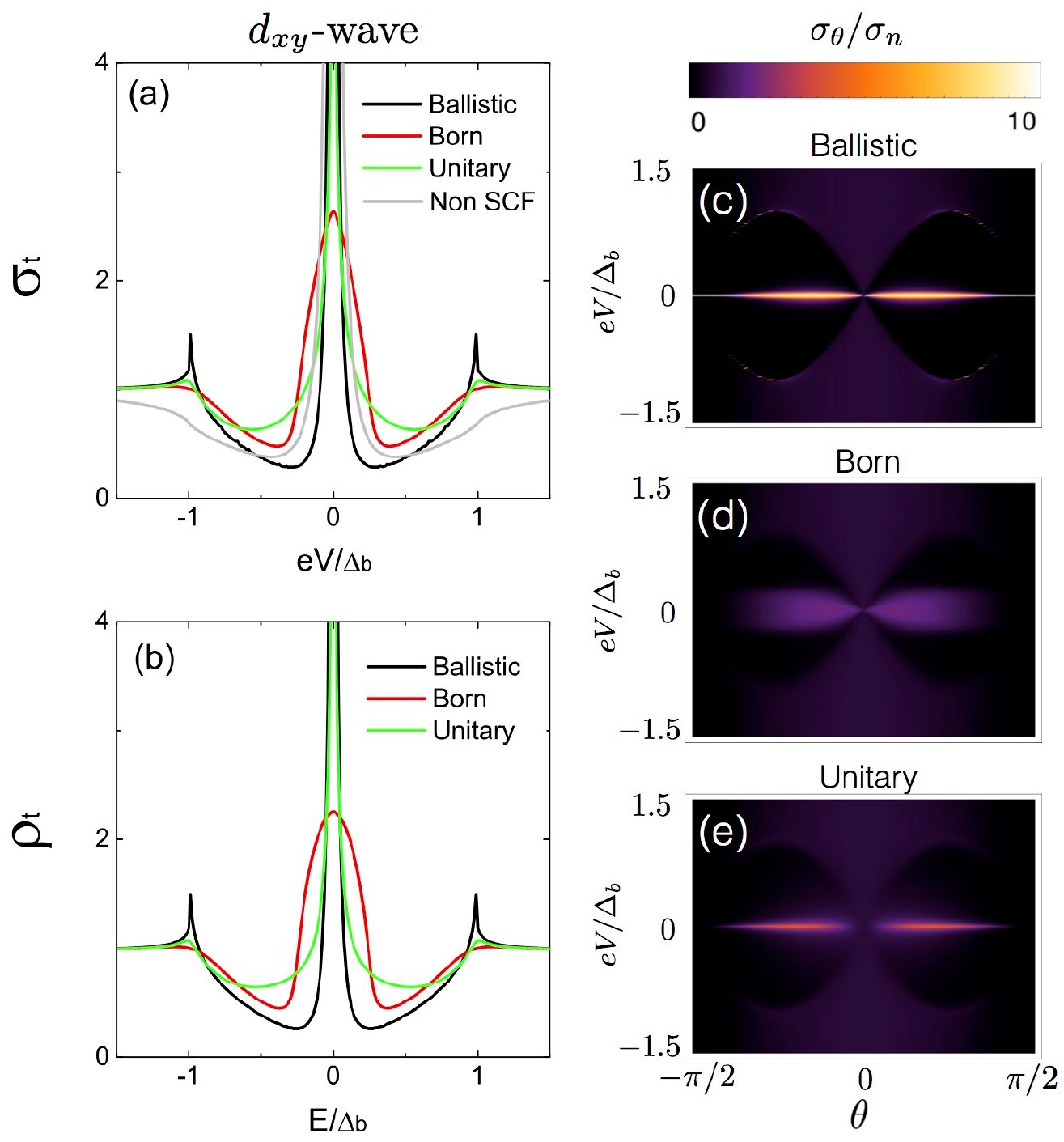}
	\caption{(Color online) $D_{xy}$-wave: (a) conductance and (b) local density of
		states at the normal metal-superconductor interface. Angle-resolved
		conductance for (c) ballistic, (d) Born, and (e) unitary limit. }
	\label{fig3}
\end{figure}

\subsection{Tunneling spectroscopy}

\begin{figure}[tbph]
	\includegraphics[width = 70 mm]{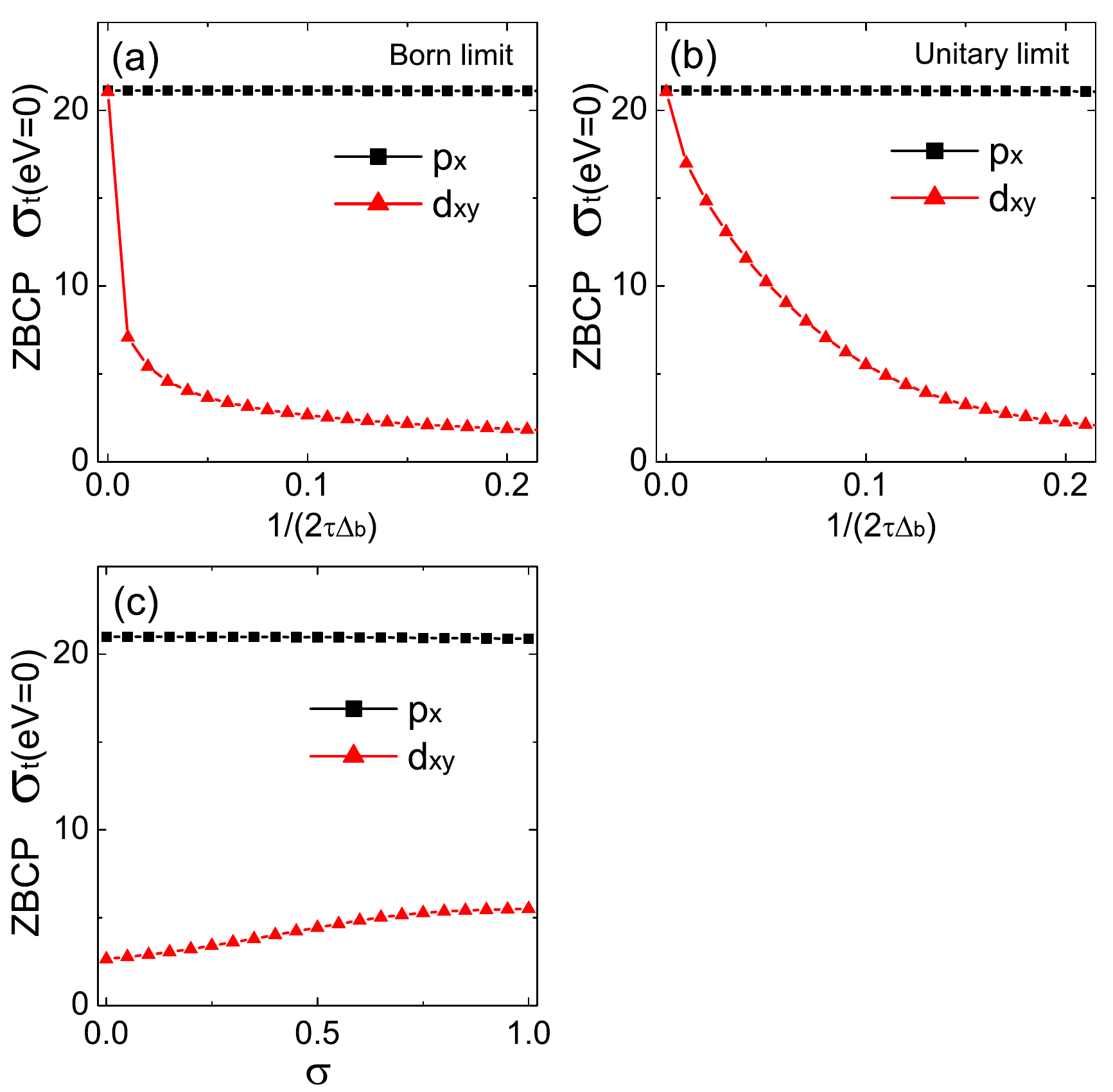}
	\caption{(Color online) Zero bias conductance peak vs. impurity scattering.
		(a) In the Born limit, the ZBCP as a function of the density of impurities and
		(b) the same plot in the unitary limit. (c) Fixing $1/\left( 2\protect\tau %
		\Delta_{b} \right) $ at $0.1$, the ZBCP as a function of the impurity strength $%
		\protect\sigma $. }
	\label{fig4}
\end{figure}

\begin{figure}[tbph]
	\includegraphics[width = 75 mm]{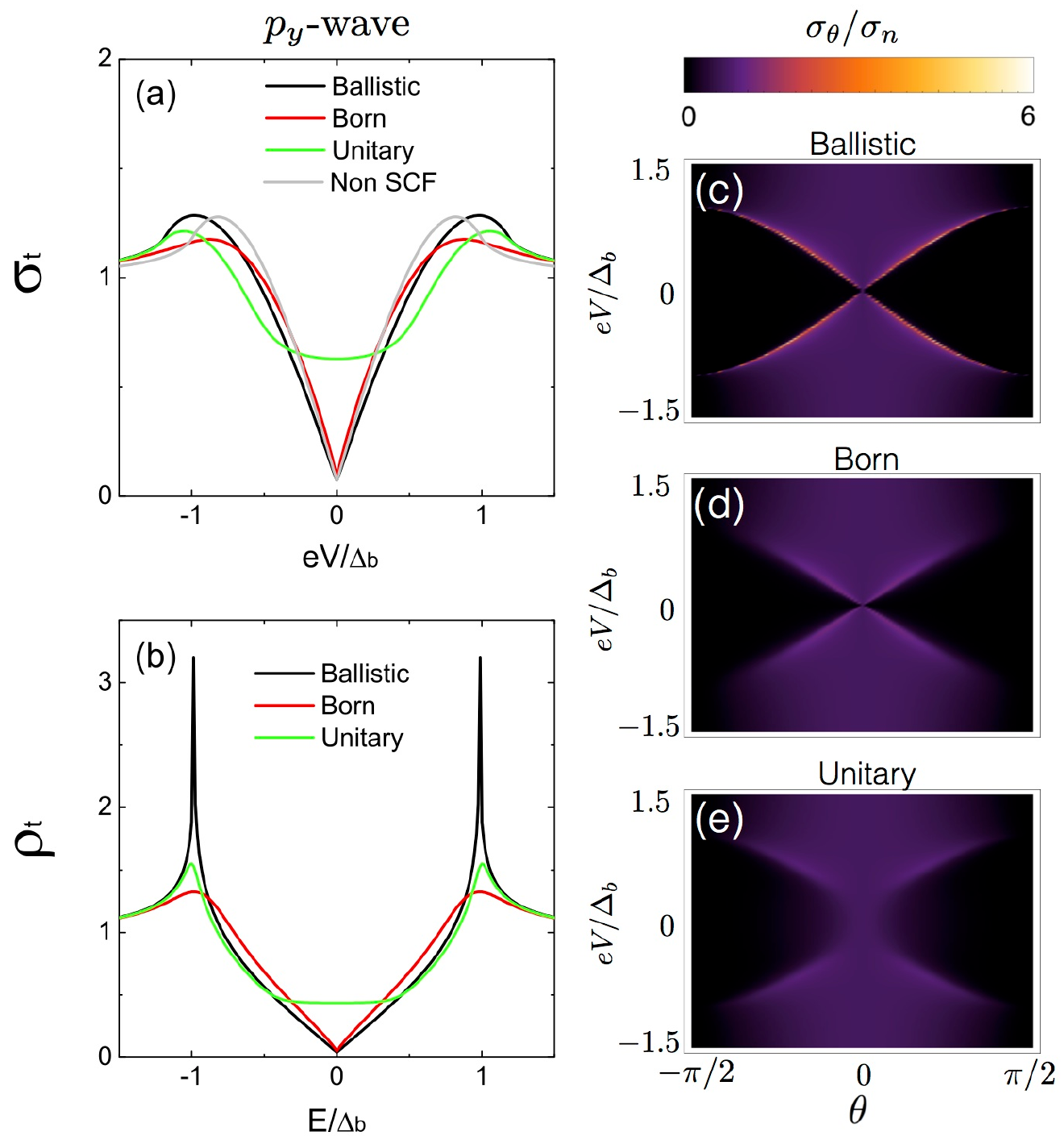}
	\caption{(Color online) $P_{y}$-wave: (a) conductance and (b) local density of
		states at the normal metal-superconductor interface. Angle-resolved
		conductance for (c) ballistic, (d) Born, and (e) unitary limit. }
	\label{fig5}
\end{figure}

We plot the normalized conductance $\sigma _{t}$ for spin-triplet $p_{x}$%
-wave in Fig.\ref{fig2}. For comparison, we also show the
non-self-consistent (non-SCF) result (light gray line) and the LDOS $\rho
_{t}=\rho _{S}/\rho _{n}$ at the interface $x=0$, with $\rho _{n}$ the
density of states in the non-superconducting state. Although the $p_{x}$%
-wave superconductor has not been found in actual materials, it is becoming a
topic of interest as a prototype for topological superconductivity\cite%
{Schnyder2008,Hasan10,Ryu10,Hasan11,ZhangRev,AliceaRev,TanakaRev2012,sato1,sato2}
and an increasing effort is dedicated to designing $p_{x}$-wave
superconductors using materials with strong spin-orbit coupling \cite%
{SauPRL,Lutchyn2010,Oreg2010,You,Ikegaya}.
Our
numerical calculation verifies the results of Ref.~\onlinecite{Ikegaya}
about the robustness of the sharp ZBCP against impurities. Furthermore, we
find that not only the height of ZBCP is robust but also the width does not
broaden. For $eV\approx \Delta _{b}$, the ballistic case shows a sharp
coherence peak which is smeared away in the presence of impurities. This is
due to the energy level broadening by impurity effect as concluded from the
angle-resolved conductance spectra $\sigma _{\theta }/\sigma _{n}$ plotted
in Figs.\ref{fig2}(d)(e). For comparison, we reproduce the conductance
result of Ref.~\onlinecite{TanumaImpurity2001} for $d_{xy}$-wave
superconductors in Fig.\ref{fig3}. Although in the ballistic limit both $%
p_{x}$- and $d_{xy}$-wave superconductors show a sharp ZBCP, their
conductance spectra under impurity scattering are quite different. The ZBCP
for $d_{xy}$-wave broadens and is suppressed by impurities, specially in the
Born limit. Figs.~\ref{fig3}(d)(e) show a smeared low energy conductance in
the angle-resolved conductance spectra compared to the ballistic case of
Fig.~\ref{fig3}(c).
The impact of impurities on the ZBCP is shown in Fig.~\ref{fig4} for $p_x$-
(black squares) and $d_{xy}$-wave (red triangles). Figs.~\ref{fig4}(a)(b)
show the evolution of the ZBCP as a function of the impurity scattering rate
in the Born and unitary limits, respectively. The ZBCP for $p_{x}$-wave
pairing is unaffected by changes in the scattering rate or the impurity
strength, shown in Fig.~\ref{fig4}(c). On the other hand, the ZBCP for $%
d_{xy}$-wave pairing is very vulnerable to the presence of impurities,
becoming rapidly suppressed. Next, we show the results for $p_{y}$-wave
junctions in Fig.\ref{fig5}. In this case, the Andreev bound state is
absent, resulting in a sharp zero bias conductance dip in the ballistic limit.
For a fixed impurity density level, we find that the dip structure is more
pronounced in the Born limit while it is greatly enhanced and smeared in the
unitary limit.

These three typical behaviors of zero bias conductance in nodal
superconductors can be well interpreted from the symmetry of odd-frequency
pairings and the divergent behavior of the self-energy. In the bulk, the
Green's function $\hat{G}\left( E,\theta \right) =\sum\nolimits_{\alpha }%
\hat{g}^{\alpha \alpha }\left( E,\theta _{\alpha }\right) $ is only
divergent when the energy locates at continuum levels. However, in the
junction system, the spacial dependent Green's function $\hat{G}\left(
E,x,\theta \right) $ can also develop a divergence near the surface (or
interface) $x=0$ due to the emergent Andreev bound state at $E=E_{b}$, $%
\theta =\theta _{b}$. For surface flat bands forming at angles connecting
gap nodes, according to Eq.\ref{eqself}, the magnitude of the components $%
a_{i}$ of the self-energy $\hat{a}\left( E=0,x=0\right) $ in the Born limit
is expected to surpass that of their counterparts in the unitary limit for
the same scattering rate $1/\tau $. 
Consequently, the ZBCP should be more
sensitive to impurities in the Born limit than in the unitary limit,
explaining the behavior of the $d_{xy}$-wave case. Further, we can also
consider the symmetry of Cooper pairs near the interface. As previously
discussed in Fig.\ref{fig1}, Cooper pairs at the interface of $p_{x}$-wave
and $d_{xy}$-wave junctions are odd in frequency. For $p_{x}$-wave,
Fermi-Dirac statistics impose that Cooper pairs form a spin-triplet
isotropic $s$-wave state, making the pairings, together with the ZBCP,
insensitive to impurities. This is a consequence of the ZBCP for $p_{x}$%
-wave junctions being protected by chiral symmetry \cite{Ikegaya}. For
pairing symmetries like $d_{xy}$-wave, Cooper pairs form an odd-frequency
spin-singlet $p$-wave pairing state which is dominant near the surface. 
Due
to the anisotropy of the state, the ZBCP is more fragile.
For pairing
symmetries supporting no Andreev bound states, like the $p_{y}$-wave case, $%
\langle \hat{G}\left( E,x=0,\theta \right) \rangle _{\theta }$ is
approximately zero when the 
\begin{figure}[tbph]
	\includegraphics[width = 75 mm]{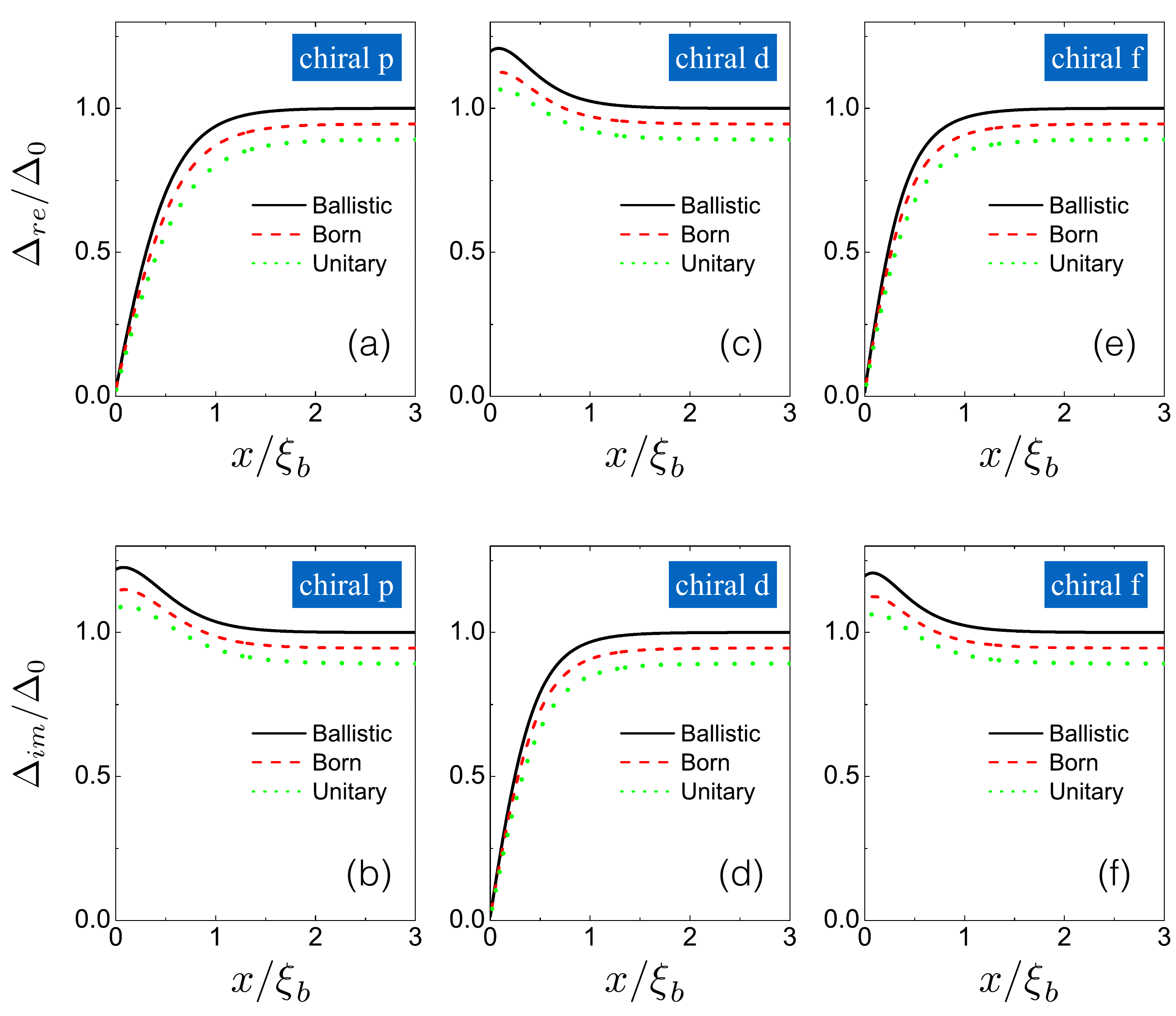}
	\caption{(Color online) Spacial dependence of order parameters in chiral
		wave superconductors: (a)(b): chiral $p$-wave, (c)(d) chiral $d$-wave, and
		(e)(f) chiral $f$-wave. The upper panels show the real part of the order
		parameter and the lower ones the imaginary part. }
	\label{fig7}
\end{figure} 
energy $E$ is away from the continuum levels,
which makes the self-energy $a_{i}$ more stronger in the unitary limit than
in the Born limit. Thus, the conductance is more sensitive in the unitary
limit as it is shown in Fig.\ref{fig5}.

\section{Gapful Superconductors with chiral surface bands}

In this section, we show the results of superconductors with chiral edge
states. There is an increasing amount of evidence supporting that Sr$_{2}$RuO%
$_{4}$ is a candidate for a chiral $p$-wave superconductor\cite%
{Maeno1994,Kallin1999,Maeno2003,Maeno2013}. The gap parameter has the form $%
\mathbf{d\propto }\left( k_{x}+ik_{y}\right) ^{\lambda }\mathbf{\hat{e}}_{z}$%
, where $\mathbf{\hat{e}}_{z}$ is a unit vector along the tetragonal crystal
$c$-axis and $\lambda =1$.
However, such a chiral triplet superconductor features a spontaneous edge
supercurrent which has not yet been observed in experiments. Several
theories have suggested that $\lambda $ could be larger than $1$, e.g., $%
\lambda =2$ (chiral $d$-wave) or $\lambda =3$ (chiral $f$-wave), leading to
a suppression of the edge current \cite{Huang14,Tada15}. Consequently, it is
crucial to study the transport signatures of chiral $d$- or chiral $f$-wave
states in Sr$_{2}$RuO$_{4}$. The study of chiral $d$-wave superconducting
states is also relevant since it has also been proposed in other systems,
e.g., in doped-graphene \cite{Schaffer14,Nandkishore12,Wang,Kiesel12}.
\begin{figure}[tbph]
	\includegraphics[width = 75 mm]{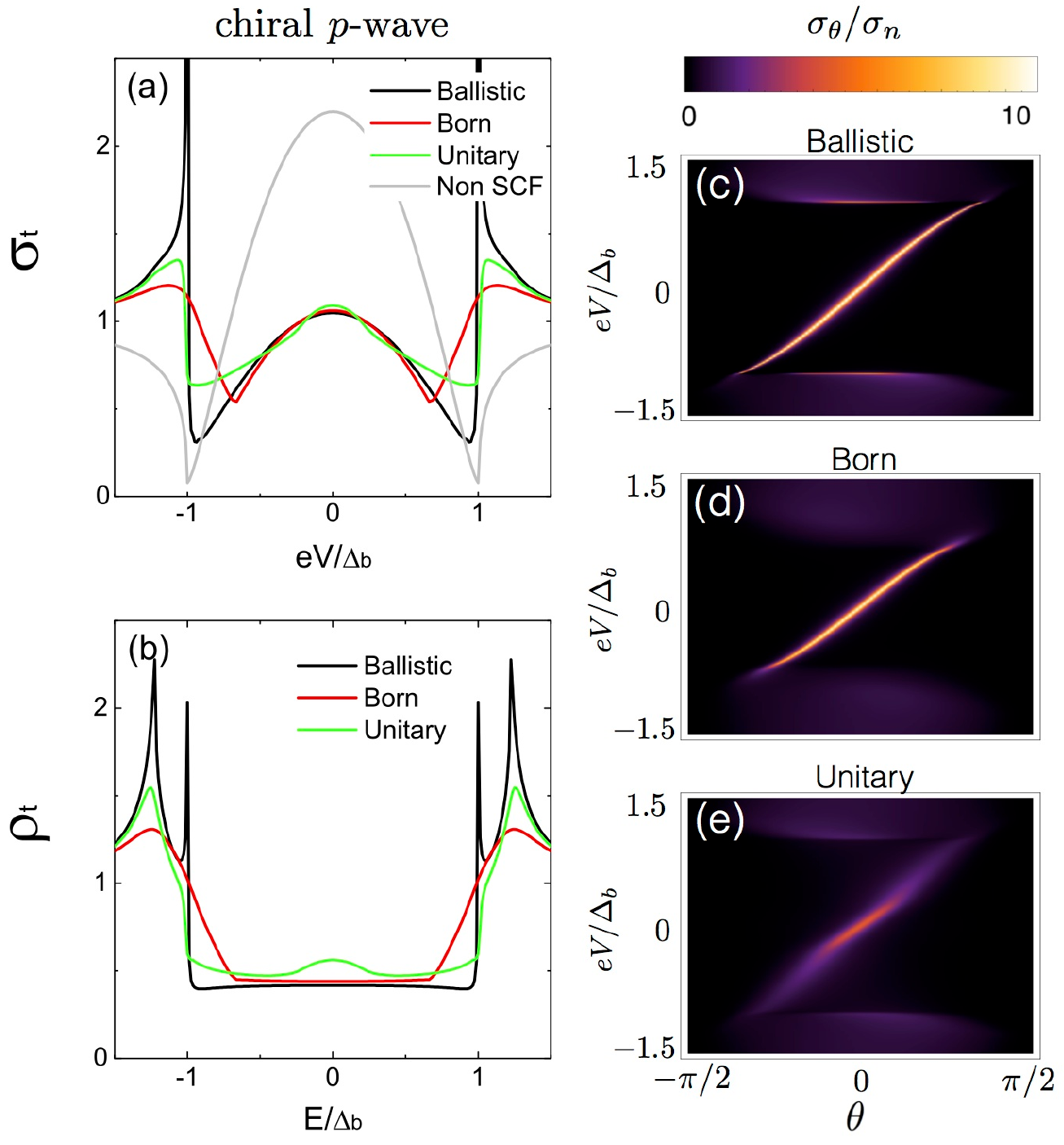}
	\caption{(Color online) Chiral $p$-wave: (a) conductance and (b) local density of
		states at the normal metal-superconductor interface. Angle-resolved
		conductance for (c) ballistic, (d) Born, and (e) unitary limit. }
	\label{fig8}
\end{figure}
\begin{figure}[tbph]
	\includegraphics[width = 75 mm]{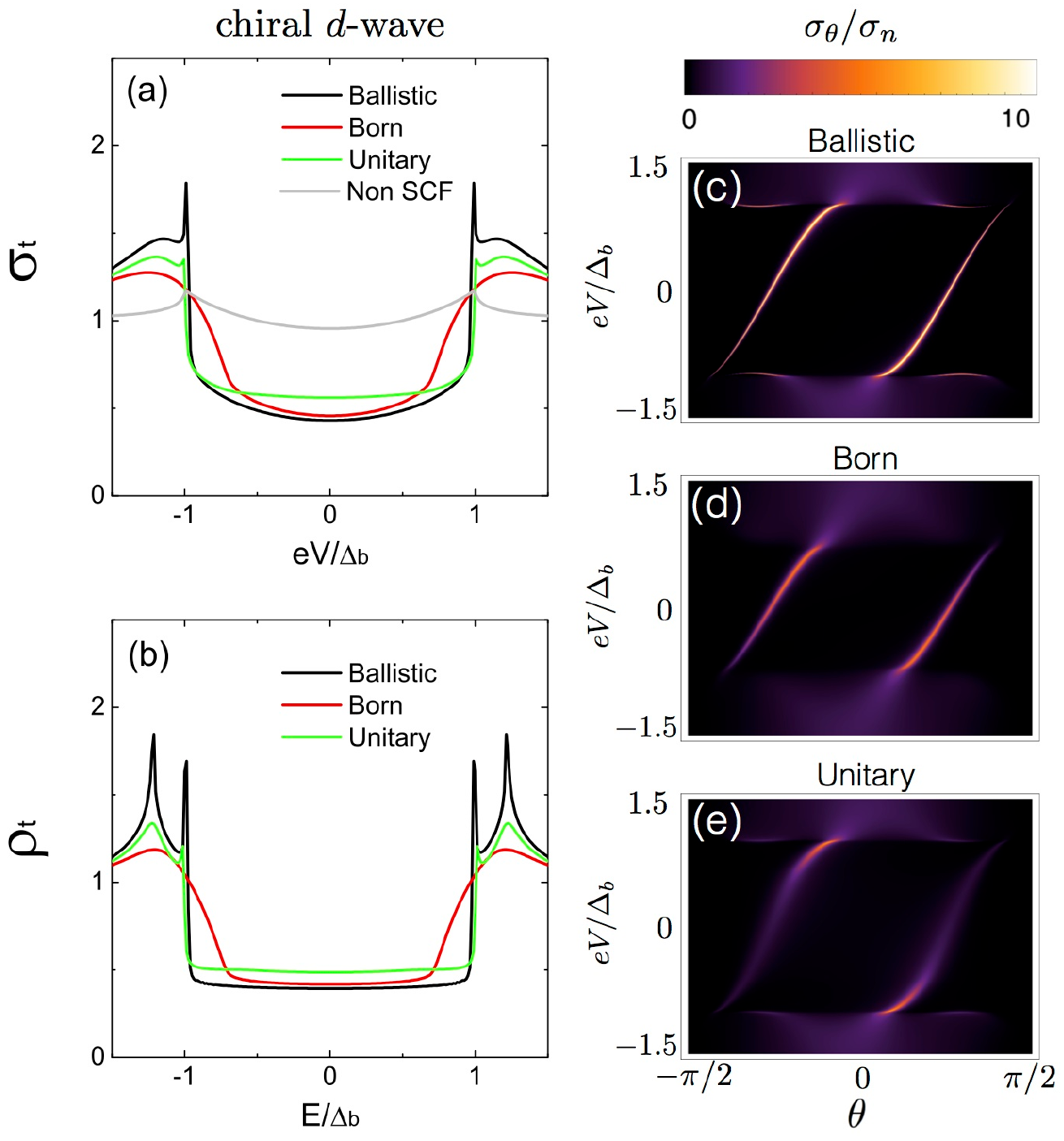}
	\caption{(Color online) Chiral $d$-wave: (a) conductance and (b) local density of
		states at the normal metal-superconductor interface. Angle-resolved
		conductance for (c) ballistic, (d) Born, and (e) unitary limit.}
	\label{fig9}
\end{figure}

First, we plot the spacial variance of order parameters in Fig.\ref{fig7}.
Chiral $p$-, $d$-, and $f$-waves display a similar behavior. The real and
imaginary components of the order parameter vary in a different way as they
approach the interface: one is slightly enhanced (real part for chiral $d$%
-wave and imaginary for chiral $p$- and $f$-wave), to a value denoted by $%
\Delta _{in}$ at the interface, and the other is greatly suppressed. Such
results have been revealed in previous studies of the BW phase of $^{3}$He%
\cite{Buchholtz1981,Higashitani2009} or Sr$_{2}$RuO$_{4}$ \cite%
{Matsumoto1999,Tanuma06,BakurskiyPRB2014}.

We now proceed to discuss the conductance and LDOS. In Fig.\ref{fig8}, we
show our results for chiral $p$-wave junctions. Compared to the non-SCF case
(gray line), the dome-like ZBCP\cite{Yamashiro97,Honerkamp,yamashira99} is still visible with or without impurity
scattering, although reduced. Actually, the height of the ZBCP has a very
weak dependence on the impurity potential. A conductance peak appears near
the bulk gap $eV\approx \Delta _{b}$, especially sharp in the ballistic
limit. Also in the ballistic case, there are double peaks in the LDOS
located at the position of $\Delta _{b}$ and $\Delta _{in}$ [see Fig. \ref%
{fig7}(b)] \cite{BakurskiyPRB2014}. The subgap LDOS at the interface is
finite and almost constant, just slightly convex, revealing the existing
chiral Andreev bound state across the region $-\Delta _{b}<E<\Delta _{b}$.
We note that although the value of the ZBCP and the subgap conductance
qualitatively agrees with the experiments, the obtained SCF result does not
reproduce the full tunneling spectra in Sr$_{2}$RuO$_{4}$\cite%
{Laube2000,Kashiwaya2011}. Thus, we consider that the self-consistent
calculation in the framework of 3-bands model\cite{Maeno2003} in Sr$_{2}$RuO$%
_{4}$ is needed in the future.

Fig.~\ref{fig9} shows the LDOS and conductance spectroscopy for chiral $d$%
-wave superconductors. As can be seen from the LDOS in panel (b), the double
peak structure outside the gap in the ballistic limit is prominent and
still visible for the unitary case at $1/\left( 2\tau \Delta _{b}\right)
=0.1$, while the peak at $E=\Delta _{b}$ is absent in the Born limit. For
the conductance spectroscopy shown in panel (a), we emphasize the absence of
ZBCP. The conductance features a wide concave bottom gap and its minimum is
barely affected by the impurity scattering. Besides that, the double peak
structure outside the gap is more visible in the conductance spectroscopy in
the ballistic and unitary limits. The angle-resolved conductance spectra in
Fig.~\ref{fig9}(c)(d)(e) display double chiral states owing to a Chern
number $ch=2$ \cite{Volovik,KKSMT}.

\begin{figure}[tbph]
	\includegraphics[width = 75 mm]{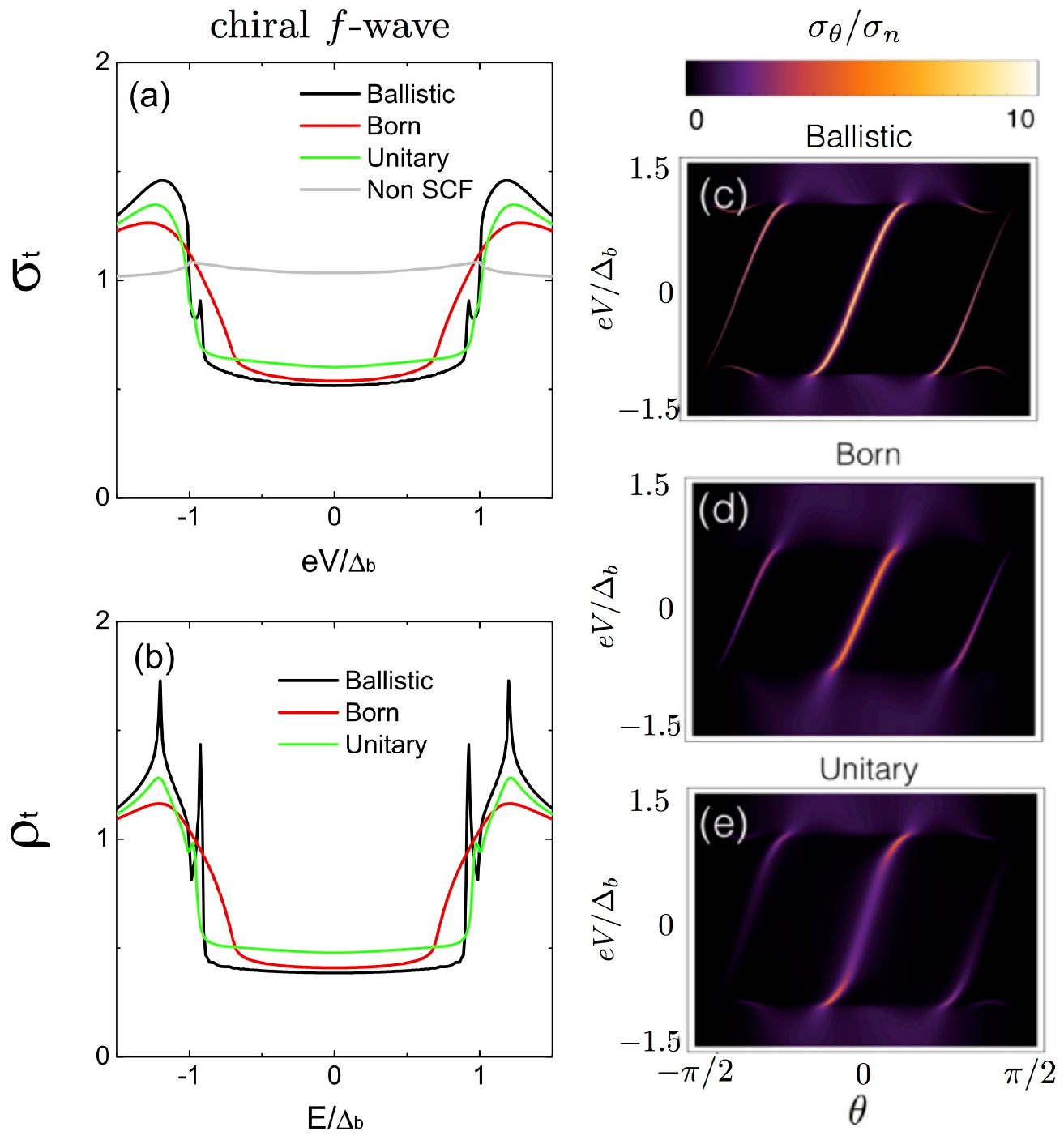}
	\caption{(Color online) Chiral $f$-wave: (a) conductance and (b) local density of
		states at the normal metal-superconductor interface. Angle-resolved
		conductance for (c) ballistic, (d) Born, and (e) unitary limit.}
	\label{fig10}
\end{figure}

Finally, we show the results for chiral $f$-wave states in Fig.\ref{fig10}.
The subgap conductance and LDOS are similar to chiral $d$-wave states as
seen in panels (a) and (b). In the ballistic limit, the conductance peak at $%
eV\approx \Delta _{b}$ becomes small as compared to chiral $p$- and chiral $d
$-wave case. The impurity effect does not change the bottom-like
characteristic of conductance and LDOS inside the bulk gap. Here, the Chern
number $ch=3$ manifests three chiral Andreev bound states inside the gap as
seen in Fig.~\ref{fig10}(c)(d)(e). However, the slopes of the states are
much higher than those of chiral $p$- and chiral $d$-wave pairings, reducing
their contribution to the angle-averaged conductance. As a result, there are
no ZBCP for the parameters we have used.

Due to the characteristics of the tunneling spectroscopy and LDOS for chiral
superconductors, it is convenient to discuss the impurity effect in three
separated energy regions: (1) $E\sim 0$; (2) $E\sim \Delta _{b}$; (3) $E\sim
\Delta _{in}$. Similarly to the previous section, let us focus on the
divergence of the Green's function $\hat{G}\left( E,\theta \right) $.
Compared to the flat surface bands, the density of zero energy states is
greatly reduced leading to a finite but relatively small value of
angle-averaged Green's function $\langle \hat{G}\left( E=0,x=0,\theta
\right) \rangle _{\theta }$. Thus, for the same scattering rate $1/\tau $,
the magnitude of the components $a_{i}$ of the self-energy $\hat{a}\left(
E=0,x=0\right) $ is more pronounced in the unitary limit than in the Born
limit. Consequently, the zero bias conductance and LDOS are more sensitive
in the unitary limit. For $E\sim \Delta _{b}$ and $E\sim \Delta _{in}$, the
appearance of continuum bands greatly enhances $\langle \hat{G}\left(
E,x=0,\theta \right) \rangle _{\theta }$ and makes the self-energy terms $%
a_{i}$ smaller. As a result, the unitary limit would retain the double peak
structure mostly outside the bulk gap $\Delta _{b}$ (e.g., see panel (b) of Fig.\ref{fig9}). It is also easy to understand that the
steep structure in the conductance and LDOS near $E\sim \Delta _{b}$ in the
ballistic limit disappears easily in the Born limit. We have summarized the
behavior of zero bias conductance in Table \ref{table}.

\begin{table}[t]
	\begin{tabular}{c||c|c|c}
		\hline
		& ballistic [$\sigma_t(eV=0)$] & Born & unitary \\ \hline\hline
		$p_x$ &  peak ($\sigma_t \gg 1$) & $\bigcirc$ & $\bigcirc$  \\ \hline
		$p_y$ &  dip ($\sigma_t=0$) & $\simeq$  & $\sim$  \\ \hline
		$d_{xy}$ &  peak ($\sigma_t \gg 1$) & - \par - & - \\ \hline
		chiral $p$ & peak ($\sigma_t \sim 1$) & $\simeq$ & $\sim$ \\ \hline
		chiral $d$ & enhancement ($0<\sigma _{t}<1$)  & $\simeq$ &  $\sim$ \\ \hline
		chiral $f$ & enhancement ($0<\sigma _{t}<1$) & $\simeq$ &  $\sim$  \\ \hline
	\end{tabular}%
	\caption{Zero bias conductance vs. impurity scattering in the ballistic, Born and unitary limit. The label $\bigcirc$ indicates the immunity to impurity scattering. $\simeq$ ($\sim$) means the zero bias conductance is approximately similar (slightly increased). $-$ ($--$) means a decrease (strong decrease) of conductance with respect to the ballistic limit. }
	\label{table}
\end{table}

\section{Conclusion}
In conclusion, we have theoretically studied the tunneling spectroscopy in normal metal-disordered unconventional superconductor junctions at low transparency for various pairing symmetries. 
For nodal superconductors with flat surface bands, the influence of the impurity scattering can be classified into three types according to the behavior of zero bias conductance: \begin{inparaenum}[(1)] 
	\item when Cooper pairs at the surface have odd-frequency $s$-wave symmetry, the zero-bias conductance peak is immune to the impurity scattering; 
	\item for non $s$-wave odd-frequency pairing, the zero-bias conductance peak is strongly suppressed by impurities; and 
	\item in the accidental absence of Andreev bound states, the zero bias conductance dip is easily smeared by impurities in the unitary limit. 
\end{inparaenum}

For chiral superconductors, we have shown the self-consistent result of conductance spectra and found that a zero-bias conductance peak for chiral $p$-wave pairings remains in the presence of impurities. The zero bias conductance for chiral waves is more sensitive to impurity scattering in the unitary limit than that in the Born limit. Moreover, the double peaks occurring in the LDOS and conductance spectra at the gap edges for all chiral waves are very sensitive to impurity scattering and are especially blurred in the Born limit. 
We expect that our results can be exploited for the experimental identification of pairing symmetries by analyzing the tunneling spectroscopy of unconventional superconductors.

\acknowledgements
We thank M. Y. Kupriyanov, K. Yada, and S. V. Bakurskiy for valuable discussions. This work was supported
by a Grant-in-Aid for Scientific Research on Innovative
Areas Topological Material Science (Grant No.
15H05853), a Grant-in-Aid for Scientific Research B (Grant
No. 15H03686), a Grant-in-Aid for Challenging Exploratory
Research (Grant No. 15K13498) from the Ministry of Education,
Culture, Sports, Science, and Technology, Japan
(MEXT); Japan-RFBR JSPS Bilateral Joint
Research Projects/Seminars (Grants No. 15-52-50054 and
No. 15668956); and Dutch FOM, the Ministry of Education and Science of the Russian Federation, Grant No. 14.Y26.31.0007
and by the Russian Science Foundation, Grant No. 15-12-30030. P.B. acknowledges support from JSPS International Research Fellowship.

%

\end{document}